\documentclass[aps,prl,twocolumn,superscriptaddress]{revtex4-2} 
\usepackage{amsmath,amssymb,graphicx,bm}
\usepackage{subcaption}
\usepackage[colorlinks=true,linkcolor=blue,citecolor=blue,urlcolor=blue]{hyperref}

\begin{document}

\title{\textit{PY-BerryAHC}: An \textit{ab-initio} python 3 code to calculate Berry Curvature dependent Anomalous Hall Conductivity in any material}

\author{Vivek Pandey}
\altaffiliation{vivek6422763@gmail.com}
\affiliation{School of Physical Sciences, Indian Institute of Technology Mandi, Kamand - 175075, India}
\author{Sudhir K. Pandey}
\altaffiliation{sudhir@iitmandi.ac.in}
\affiliation{School of Mechanical and Materials Engineering, Indian Institute of Technology Mandi, Kamand - 175075, India}

\date{\today}

\begin{abstract}

The anomalous Hall conductivity (AHC) in materials has long been a topic of debate. Recent studies have established that AHC originates from the Berry curvature ($\boldsymbol\Omega$) of Bloch states. Accurate computation of AHC is crucial for predicting material properties, validating theoretical models, and guiding experimental studies in topological and spintronic applications. Traditional approaches often rely on wannier interpolation, which can introduce inaccuracies and computational overhead. Additionally, reliability of the wannierization technique becomes questionable when the bands in the region of interest are highly entangled and dispersive. This demands the calculation of AHC using the \textit{first-principle} approach. In this work, we present \textit{PY-BerryAHC}, a Python 3 based code that directly computes $\boldsymbol\Omega$ and AHC using WIEN2k output, eliminating the need for wannierization. Since, WIEN2k employs an all-electron full-potential linearized augmented plane wave (FP-LAPW) method, \textit{PY-BerryAHC} provides highly accurate AHC results. Furthermore, the code efficiently handles large \textbf{\textit{k}}-grids by parallelizing $\boldsymbol\Omega$ computations over \textbf{\textit{k}}-points. Additionally, it stores band-resolved $\boldsymbol\Omega$ in a binary file, thereby greatly reducing the required storage memory and allowing fast post-processing to compute AHC for a range of temperature ($T$) and chemical potential ($\mu$) values without re-evaluating $\boldsymbol\Omega$. \textit{PY-BerryAHC} has been validated on well-known materials exhibiting AHC. These include- Fe, Fe$_3$Ge \& Co$_2$FeAl. At 300 K, the calculated magnitude of $\sigma_{xy}$ for Fe \& Fe$_3$Ge is found to be 744 $S/cm$ \& 311 $S/cm$, respectively. For Co$_2$FeAl, the magnitude of $\sigma_{xy}$ is obtained to be $\sim$56 $S/cm$ and is found to be constant with the change in temperature from 0-300 K. These results are in good agreement with previously reported theoretical and experimental data. This ensures the accuracy, reliability and efficiency of the code. Additionally, the automated workflow of the code, requiring no manual tuning, makes it a robust and efficient tool for high-throughput studies of topological materials. Apart from these features, the code is also provided with a post-processing tool named \textit{berry\_plot.py} which generates an interactive window with various options to visualize the effective $\boldsymbol\Omega$ [$\Omega_\xi(\textbf{\textit{k}})$=$\sum_n f_n(\textbf{\textit{k}}) \Omega_\xi^n(\textbf{\textit{k}})$ ($\xi$=$x$, $y$ or $z$)] across the Brillouin zone (BZ). This helps in better understanding of the origin of AHC in any material.\\\\

\textbf{Program summary -}\\
\textit{Program Title}: \textit{PY-BerryAHC}\\
\textit{Program Files doi}:\\
\textit{Licensing provisions}: GNU General Public License 3.0\\
\textit{Programming language}: Python 3\\
\textit{External routines/libraries}: Math, Time, Numpy, Multiprocessing, Uuid, Struct\\
\textit{Nature of problem}: Calculating AHC of any material using the \textit{first-principle} approach.\\
\textit{Solution method}: The code calculates the anomalous Hall conductivity (AHC) by first computing the $\boldsymbol\Omega$ using the Kubo formula. For this, the eigenvalues and momentum-matrices are obtained directly from the \textit{first-principle} calculations using WIEN2k. $\boldsymbol\Omega$ is then integrated over the entire BZ, weighted by the Fermi-Dirac distribution function to account for the electronic occupation at different chemical potentials and temperatures.
\end{abstract}

\maketitle

\section{Introduction}
The anomalous Hall conductivity (AHC) is a fundamental transport property of materials in which time-reversal symmetry is broken. AHC arises from the Berry curvature ($\boldsymbol\Omega$) of electronic bands in momentum space\cite{hall1880rotational,nagaosa2010anomalous,nayak2016large}. Unlike the conventional Hall effect, which originates from Lorentz forces acting on charge carriers\cite{yao2024geometrical}, the anomalous Hall effect (AHE) is an intrinsic quantum mechanical phenomenon governed by the topology of the electronic structure\cite{weng2015quantum}. $\boldsymbol\Omega$ acts as an effective magnetic field in \textbf{\textit{k}}-space\cite{nagaosa2012emergent}, playing a crucial role in various topological materials, including Weyl semimetals (WSMs)\cite{burkov2014anomalous,pandey2021anab}, magnetic Weyl systems\cite{meng2019large}, and quantum anomalous Hall insulators\cite{ferguson2023direct}. Studies reveal that AHC in a material can greatly enhance its practical applicability for various purposes. Some of them are discussed further.

AHC attributed from non vanishing $\boldsymbol\Omega$ in a material leads to dissipationless Hall currents which is very useful for spintronics application\cite{fukami2020antiferromagnetic,miah2007observation} and quantum topological phenomena\cite{bradlyn2017topological,wieder2020magnetic}. Apart from this, such intrinsic AHC has been found to have numerous applications in the diluted magnetic semiconductors (DMS)\cite{PhysRevLett.88.207208}. Moving further, a recent experimental work reveal that symmetry elements of Heusler magnets can be changed such that the $\boldsymbol\Omega$ and all the associated properties are switched while leaving the magnetization unaffected. This helps in tuning the AHC from 0 to $\sim$1600 $Ohm^{-1}cm^{-1}$ in full Heusler alloys Co$_2$MnGa \& Mn$_2$CoGa\cite{manna2018colossal}. Such tunability of AHC is expected to enhance the practical applications of these compounds. These discussions highlight the importance of AHC in engineering technological applications. Computational methods have always played a vital role in investigating various properties of a given material. In this regard, accurate computation of $\boldsymbol\Omega$ and AHC will prove to be an effective tool for better understanding of electronic transport in a given material and for designing novel topological devices.

Theoretically, AHC is computed by integrating $\boldsymbol\Omega$ over all the bands at each \textbf{\textit{k}}-points across the Brillouin zone (BZ), with contributions weighted by the Fermi-Dirac distribution to reflect electronic occupation at a given chemical potential ($\mu$) and temperature ($T$). The expression for AHC is given by: $\sigma_{x y}^{\mathrm{AHC}}=-\frac{e^2}{\hbar} \frac{1}{(2 \pi)^3} \sum_n \int_{\mathrm{BZ}} f_n(\textbf{\textit{k}}) \Omega_z^n(\textbf{\textit{k}}) d \textbf{\textit{k}}$, where $f_n(\textbf{\textit{k}})$ is the Fermi-Dirac distribution function and $\Omega_z^n(\textbf{\textit{k}})$ is the $z$-component of $\boldsymbol\Omega$ for band-index $n$ and at crystal momentum \textbf{\textit{k}}\cite{PhysRevB.85.012405,xiao2010berry}. From the well-known Kubo formula for computing $\Omega_z^n(\textbf{\textit{k}})$, it is known that the term is highly sensitive to the details of electronic structure of the material under study, specially on the energy eigenvalue and the momentum-matrices associated with the particular (\textit{n},\textbf{\textit{k}}) state\cite{chang1996berry,sundaram1999wave,onoda2002topological,jungwirth2002anomalous}. Thus, the accuracy of the calculated AHC depends on how precise and accurate computation of energy eigenvalue and the momentum-matrices of (\textit{n},\textbf{\textit{k}}) state is done. This in turn depends on how accurately the (\textit{n},\textbf{\textit{k}}) is determined. Moreover, an accurate and efficient numerical approach is required to evaluate the integral over a dense \textbf{\textit{k}}-point grid.

Presently, the most common approach to calculate AHC is based on tight-binding (TB) model. Such TB models are usually generated using the wannierization techniques as implemented in Wannier90 package\cite{mostofi2008wannier90}. In this method, maximally localized Wannier functions (MLWFs) are constructed from the Bloch states obtained in density functional theory (DFT) calculations\cite{marzari2012maximally,marrazzo2024wannier}. These MLWFs serve as basis set, in terms of which, Bloch states (\textit{n},\textbf{\textit{k}}) corresponding to any band-index \textit{n} \& crystal momentum \textit{\textbf{k}} is reproduced using wannier interpolation techniques. Using these reproduced Bloch states, the energy eigenvalues and momentum-matrices are calculated, which are further used to compute $\Omega_z^n(\textbf{\textit{k}})$ \& AHC. This method is implemented in packages like Wannier90\cite{mostofi2008wannier90} and WannierTools\cite{wu2018wanniertools}. It is important to note here that the accuracy of calculated AHC depend on the quality of the wannier functions obtained via wannierization technique. For the case of AHC computation, the high quality wannier functions (HQWFs) are one that perfectly reproduce the Bloch states with correct energy and associated momentum-matrices in the energy-region of interest. Moreover, the number of these wannier functions must be as small as possible. However, to obtain HQWFs, the DFT bands in the required energy window must be disentangled from the bands outside this window. The window is popularly known as disentangled energy window. Additionally, the projectors contributing to the bands in disentangled energy window must have negligible contribution to other bands outside the window. Another general observation is that obtaining HQWFs becomes extremely difficult when the bands in the region of interest are greatly entangled among themselves and are highly dispersive. Apart from these, wannierization procedure is highly parameter dependent and there is no straight-forward procedure to obtain HQWFs\cite{pandey2023py}. Moving to the case of real materials, there may be the case where the dispersion curve may not possess disentangled energy window. Furthermore, the bands may be extremely dispersive as in case of metals. In such situations, obtaining HQWFs is a hopeless situation and thus the reliability of AHC obtained from the wannier interpolation becomes questionable. 

An alternative to Wannier-based methods is the direct calculation of $\boldsymbol\Omega$ from \textit{first-principles} wave functions. This approach avoids interpolation errors by working directly with the original DFT eigenfunctions and momentum-matrix elements, thereby eliminating the need for wannier functions. Such implementation has been previously carried out in the LmtART package, which uses full-potential linear muffin-tin orbital (FP-LMTO) basis set\cite{savrasov2005program}. In contrast to LmtART package, WIEN2k package uses full-potential linearized augmented-plane-wave (FP-LAPW) basis set\cite{blaha2020wien2k}. In the present DFT community, FP-LAPW basis set is claimed to be the most accurate one\cite{wimmer1981full,jansen1984total,mattheiss1986linear,BLAHA1990399}. Hence, AHC computed from a code that directly uses eigenvalues and momentum-matrices obtained from WIEN2k calculations is expected to be more accurate and reliable. Additionally, WIEN2k is one of the most widely used DFT package. Thus, implementing the calculation of AHC using the output of WIEN2k calculations will be beneficial for vast research audience. Now we shall discuss the aspects that can enhance the efficiency of code for such implementation.

The transport properties such as AHC generally get converged for \textbf{\textit{k}}-mesh size of 400$\times$400$\times$400 or more. Moreover, if the bands\textquoteright \hspace*{0.02in} resolved $\boldsymbol\Omega$ at each \textit{\textbf{k}}-points is stored, then AHC can be easily calculated at any temperature and chemical potential in the form of post-process. In such a scenario, parallel computing and efficient handling of data will greatly enhance the efficiency of the code. By efficient handling of data, we mean that one which occupy less memory and is easily accessible. This can be achieved by storing the coordinates of \textit{\textbf{k}}-points and the values of bands\textquoteright \hspace*{0.02in} resolved $\boldsymbol\Omega$ in a binary file. The binary file is not directly accessible. Thus, one can only store important information without any extra formatting or comment lines. Furthermore, storing data in binary format greatly reduces the size of the file (about 10-15 times). Moving to parallel computing, for the present case the calculation of $\boldsymbol\Omega$ can be made parallel over \textbf{\textit{k}}-points. In addition to this, the calculation of $\sum_n f_n(\textbf{\textit{k}}) \Omega_z^n(\textbf{\textit{k}})$ in the post-process for AHC computation can also be made parallel over \textbf{\textit{k}}-points. A \textit{first-principle} code designed to calculate AHC using these features is expected to be very efficient and less time consuming.

In this work, we develop a python 3 based code named \textit{PY-BerryAHC} to compute $\boldsymbol\Omega$ and AHC directly from WIEN2k output. The code extracts eigenvalues and momentum-matrix elements from WIEN2k computations and evaluates $\boldsymbol\Omega$ explicitly by invoking the Kubo-like formula\cite{thouless1982quantized}. For efficient handling of large \textit{\textbf{k}}-points, calculations of $\boldsymbol\Omega$ is made parallel over \textit{\textbf{k}}-points. Furthermore, unlike in WannierTools where calculations of $\boldsymbol\Omega$ and AHC is done corresponding to a single temperature at a time, \textit{PY-BerryAHC} first calculates band-resolved $\boldsymbol\Omega$ and stores it in a binary file (thereby reducing the memory required for storage). Then, based on the value of temperature window and chemical potential window provided in the input file, it computes AHC as post-process. While the majority of computation time is involved in calculations of $\boldsymbol\Omega$, the post-process is comparatively very fast. Thus, \textit{PY-BerryAHC} can be used to efficiently compute AHC for a range of chemical potential and temperature values in a single run once the $\boldsymbol\Omega$ file is obtained. Additionally, \textit{PY-BerryAHC} provides a direct and interpolation-free alternative, ensuring numerical accuracy and scalability. This is particularly advantageous for high-throughput materials discovery, where manual fine-tuning of Wannier functions is impractical. To validate the code, it has been tested on some well-known materials exhibiting AHC. These includes- Fe, Fe$_3$Ge and Co$_2$FeAl. The magnitude of AHC obtained using the \textit{PY-BerryAHC} code is also compared with the values reported in literatures corresponding to each material. Also, the code is equipped with a module named \textit{berry\_plot.py} which provides several interactive options to visualize the variation of $\Omega_\xi(\textbf{\textit{k}})$=$\sum_n f_n(\textbf{\textit{k}}) \Omega_\xi^n(\textbf{\textit{k}})$ ($\xi$=$x$, $y$ or $z$) across the BZ. This further enhances the capability of the code in exploring topological materials.

\begin{table*}
\caption{\label{tab:table1}%
\normalsize{The details of various input parameters for \textit{PY-BerryAHC} code.
}}
\begin{ruledtabular}
\begin{tabular}{cc}
\textrm{\textbf{Name}}&
\textrm{\textbf{Meaning}}\\
\colrule
    \textit{case} &  WIEN2k self-consistently energy converged file name (SOC must be included).\\
    \textit{struct\_num} & Crystal structure number. (Refer Table II)\\ 
    \textit{chemical\_potential} & Chemical potential (in Rydberg) around which \\
                                 & chemical-potential window is created corresponding to which AHC is computed.\\ 
    \textit{emax} & Upper energy limit (in meV) with respect to \textit{chemical\_potential} of the chemical-potential window.\\ 
    \textit{emin} & Lower energy limit (in meV) with respect to \textit{chemical\_potential} of the chemical-potential window.\\ 
    \textit{estep} & Steps (in meV) with which the chemical potential value must be changed within the chemical-potential window.\\ 
    \textit{Tmin} & Lower temperature limit (in Kelvin) of the temperature window within which AHC must be computed.\\
    \textit{Tmax} & Upper temperature limit (in Kelvin) of the temperature window within which AHC must be computed.\\
    \textit{Tstep} & Steps (in Kelvin) with which the temperature value must be changed within the temperature window.\\
    \textit{spin\_pol} (1/0) & Spin-polarization is included (1) or not (0). SOC is intrinsically included for \textit{PY-BerryAHC} calculations\\
                            & when the code is run for spin-polarized case.\\ 
    \textit{kgrid} & The size of \textbf{\textit{k}}-mesh to be produced at which $\boldsymbol\Omega$ must be computed.\\ 
    \textit{shift\_in\_k} (1/0) & Shift of \textbf{\textit{k}}-mesh from high-symmetric points is needed (1) or not (0). \\ 
    \textit{NUM\_THREADS} & Number of threads over which computation of $\boldsymbol\Omega$ and AHC over \textbf{\textit{k}}-points must be made parallel.\\ 
    \textit{num\_kp\_at\_once} & At each thread, $\boldsymbol\Omega$ at \textit{num\_kp\_at\_once} number of \textbf{\textit{k}}-points\\
                               &  will be computed at once and then the next \textit{num\_kp\_at\_once} number of \textbf{\textit{k}}-points will be\\
                               &  processed. This is to avoid the calculations from being aborted due to higher occupation of RAM.\\ 
    \textit{berry\_plot\_at\_T} & Temperature (in Kelvin) at which $\Omega_\xi(\textbf{\textit{k}})$=$\sum_n f_n(\textbf{\textit{k}}) \Omega_\xi^n(\textbf{\textit{k}})$ \\
                              & must be plotted using the \textit{berry\_plot.py} module of \textit{PY-BerryAHC} code. Here, $\xi$=$x$, $y$ or $z$.
\end{tabular}
\end{ruledtabular}
\label{tab1}
\end{table*}

\begin{table}
\caption{\label{tab:table1}%
\normalsize{The structure number assigned to different crystal structures.
}}
\begin{ruledtabular}
\begin{tabular}{cc}
\textrm{\textbf{Crystal Structure}}&
\textrm{\textbf{Structure Number}}\\
\colrule
          Cubic Primitive           & 1\\
          Cubic face-centred        & 2\\
          Cubic body-centred        & 3\\
          Tetragonal Primitive      & 4\\
          Tetragonal body-centred   & 5\\
          Hexagonal Primitive       & 6\\
          Orthorhombic Primitive    & 7\\
          Orthorhombic base-centred & 8\\
          Orthorhombic body-centred & 9\\
          Orthorhombic face-centred & 10\\
          Trigonal Primitive        & 11 \\
    
\end{tabular}
\end{ruledtabular}
\label{tab2}
\end{table}

\section{Theoretical Background}

\subsection{Berry curvature and anomalous Hall conductivity}
As previously discussed, AHC originates from the $\boldsymbol\Omega$ of electronic bands in momentum space. \textit{PY-BerryAHC} directly calculates the antisymmetric tensor $\Omega_{\mu \nu}^n$ from the WIEN2k output by invoking the Kubo-like formula\cite{stejskal2023theoretical}:
\begin{equation}
   \Omega_{\mu \nu}^n(\textbf{\textit{k}})=-\frac{\hbar^2}{m^2} \sum_{n^{\prime} \neq n} \frac{2 \operatorname{Im}\left\langle\psi_{n \textbf{\textit{k}}}\right| p_\mu\left|\psi_{n^{\prime} \textbf{\textit{k}}}\right\rangle\left\langle\psi_{n^{\prime} \textbf{\textit{k}}}\right| p_\nu\left|\psi_{n \textbf{\textit{k}}}\right\rangle}{\left(E_n-E_{n^{\prime}}\right)^2}
   \label{eq2}
\end{equation}
where $E_n$ is the band-energy, $\psi_{n \textbf{\textit{k}}}$ are the Bloch states and $p_\mu$ are the momentum operators. In addition to this, $\hbar$ \& $e$ denote the reduced Planck\textquoteright s constant and mass of free electron, respectively. In three dimensions, the $\Omega_{\xi}$ component of $\boldsymbol\Omega$ can be obtained from the  $\Omega_{\mu\nu}$ using a pseudovector relation given by $\Omega_{\mu\nu}$ = $\epsilon_{\mu\nu\xi} \Omega_{\xi}$, where, $\mu$, $\nu$ \& $\xi$ represent $x$, $y$ \& $z$ cartesian direction as per required by the expression. Using $\Omega_{\xi}$, AHC ($\sigma_{\mu \nu}^{\mathrm{AHC}}$) is calculated by using the formula\cite{yao2004first,xiao2010berry,ernst2019anomalous,helman2021anomalous}:
\begin{equation}
\sigma_{\mu \nu}^{\mathrm{AHC}}=-\frac{e^2}{\hbar} \frac{1}{(2 \pi)^3} \sum_n \int_{\mathrm{BZ}} f_n(\textbf{\textit{k}}) \Omega_\xi^n(\textbf{\textit{k}}) d \textbf{\textit{k}}
\label{eq31}
\end{equation}
where $f$ is the Fermi-Dirac distribution function. Here, $\Omega_\xi^n(\textbf{\textit{k}})$ denotes the $\Omega_{\xi}$ component of $\boldsymbol\Omega$ associated with the band with band-index \textbf{\textit{n}} at the \textbf{\textit{k}}-point \textbf{\textit{k}}. Moving further, the expression shows that AHC results from integrating the $\Omega_\xi^n(\textbf{\textit{k}})$ weighted by the occupation of states (given by $f_n(\textbf{\textit{k}})$) over the entire BZ. The presence of $f$ in equation \ref{eq31} allows the computation of AHC at various combination of chemical potential and temperature.

An important aspect to highlight here is the role of spin-orbit coupling (SOC) in determining AHC. If SOC strength in a system is zero, then the up-spin and down-spin sub systems are independent and each is governed by a real Hamiltonian. In such a scenario, $\Omega_{\xi}^n(\textbf{\textit{k}})$=$-\Omega_{\xi}^n(-\textbf{\textit{k}})$ for all combinations of $n$ \& \textbf{\textit{k}}. Thus, the integral in equation \ref{eq31} vanishes. This situation arises because the spontaneously broken time-reversal symmetry in the spin channel is not communicated to the orbital channel, which is responsible for giving rise to AHC\cite{vanderbilt2018berry}. Thus for computing AHC, SOC-effect must be necessarily taken into account. \textit{PY-BerryAHC} is designed considering this aspect.

\subsection{Extracting Berry curvature from WIEN2k output}
WIEN2k is a full-potential linearized augmented plane wave (FP-LAPW) method-based DFT package\cite{blaha2020wien2k}. The momentum-matrix elements required for $\boldsymbol\Omega$ calculations can be obtained using WIEN2k's OPTIC module, which computes: $\textless \psi_{n\textbf{k}}| \textbf{\textit{p}}| \psi_{m\textbf{k}} \textgreater$, where $\textbf{\textit{p}}$ is the momentum operator. The workflow for extracting $\boldsymbol\Omega$ involves- (I) Perform self-consistent SOC included DFT calculations to obtain ground state charge density, (II) obtain the eigen values at a given \textbf{\textit{k}}-point, (III) obtain the momentum-matrix elements at the given \textbf{\textit{k}}-point, and (IV) calculate $\Omega_{\mu \nu}^n(\textbf{\textit{k}})$ at the given \textbf{\textit{k}}-point using equation \ref{eq2}. In the present DFT community, FP-LAPW basis set is considered as the most accurate one. Thus, the obtained values of $\Omega_{\mu \nu}^n(\textbf{\textit{k}})$ are expected to be highly accurate.

\subsection{Numerical integration over the entire BZ}
It is seen in equation \ref{eq31} that $\sigma_{\mu \nu}^{\mathrm{AHC}}$ involves the integration of $f_n(\textbf{\textit{k}}) \Omega_\xi^n(\textbf{\textit{k}})$ over the entire BZ. In \textit{PY-BerryAHC}, this integral is approximated by computing the sum of $f_n(\textbf{\textit{k}}) \Omega_\xi^n(\textbf{\textit{k}})$ over a dense \textbf{\textit{k}}-mesh (uniformly distributed) across the BZ.

\section{Workﬂow and technical details}

\subsection{Workﬂow}

\begin{figure}[t]
    \centering
    \includegraphics[width=0.50\textwidth,height=14.0cm]{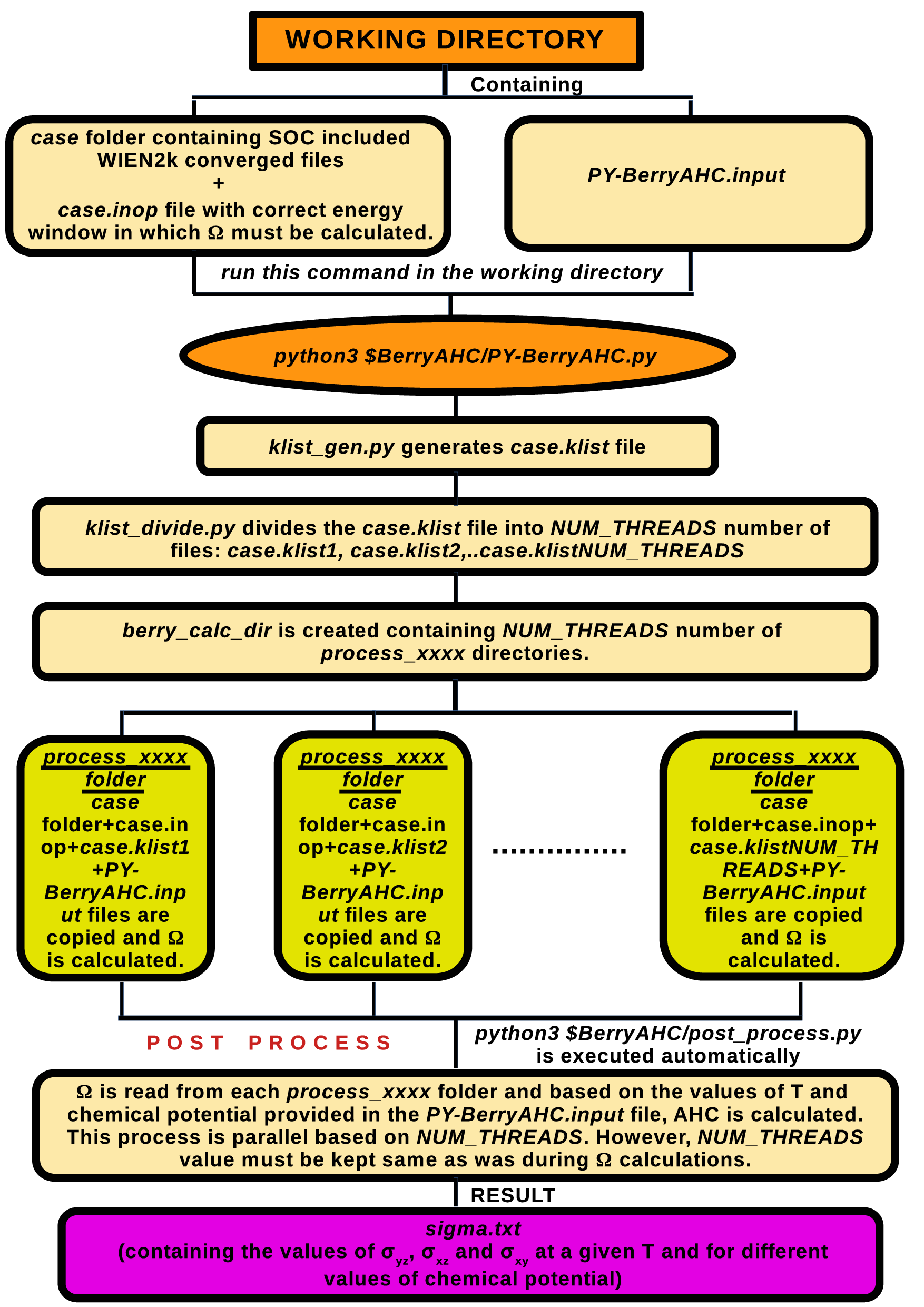}
    \caption{Workﬂow of the \textit{PY-BerryAHC} code.}
    \label{workflow}
\end{figure}

For using the \textit{PY-BerryAHC} code, user needs to install it by running \textit{setup.py} module provided with the package. The command for this is: \textit{python3 setup.py}. This will create the \textit{bin} folder containing all the necessary modules. The procedure will also define the path of this \textit{bin} folder in \textit{.bashrc} file with the variable named \textit{BerryAHC}. Now everything is set for AHC calculations using \textit{PY-BerryAHC} package. The detailed procedure for running the code is described below.

For running the AHC calculations, one needs to first obtain the \textit{case} folder containing SOC included self-consistently converged ground state files corresponding to the given material from WIEN2k package. Nextly, \textit{PY-BerryAHC.input} file must be prepared. All the necessary input variables corresponding to AHC calculations must be defined in it. These input variables are briefly described in table \ref{tab1}. The user needs to specify the name of folder containing SOC included self-consistently converged ground state files to the \textit{case} variable. Nextly, the structure number of the material under study must be specified in front of \textit{struct\_num} variable in the input file. The structure number assigned to different crystal structure is depicted in table \ref{tab2}. Moving next, \textit{PY-BerryAHC} is capable of calculating AHC in the range of chemical potential and for each of these values, it also computes AHC in range of temperature values at a time. For this, the temperature and chemical potential window must be defined. In this regard, the lower-limit, upper-limit and the temperature steps with which temperature values must be changed (in between the defined temperature limits) must be assigned to variables \textit{Tmin}, \textit{Tmax} \& \textit{Tstep}, respectively. All these values must be given in Kelvin. Here, one must note that if calculation is required for single temperature value, lets say at $T_1$, then both \textit{Tmin} and \textit{Tmax} must be set to $T_1$.  Furthermore, the chemical potential window is defined with respect to the value assigned to variable \textit{chemical\_potential} (in Rydberg). The lower-limit, upper-limit and energy-steps with which the chemical potential values must be changed within the chemical potential window must be allocated to the variables  \textit{emax}, \textit{emin} \& \textit{estep}, respectively. These variables take values in meV. Here, it is important to highlight that if \textit{emax} \& \textit{emin} are set to 0, then AHC is computed at single value of chemical potential that is assigned to \textit{chemical\_potential} variable. Moving further, if a material is magnetic, then eigenvalues and momentum-matrices must be calculated by including the spin-polarization limit. In this regard, one should specify 1 in \textit{spin\_pol} variable if spin-polarization is needed, else 0 must be assigned to the variable. One must note here that the spin-polarized calculations is performed in \textit{PY-BerryAHC} by taking into account the effect of SOC. Nextly, \textit{PY-BerryAHC} can be run in serial or parallel mode. For this, user needs to assign number of threads to be used to \textit{NUM\_THREADS} variable. If \textit{NUM\_THREADS}=1, the code will run in serial mode. It is important to mention here that \textit{PY-BerryAHC} is made parallel by using the Pool class of multiprocessing module of Python 3. Moving next, it is important to highlight here that for generating \textit{case.klist} corresponding to higher size of \textit{k}-mesh, WIEN2k takes considerably long time. Keeping this in mind, \textit{PY-BerryAHC} is provided with a separate module named \textit{klist\_gen.py} which easily generates the \textit{case.klist} file. For this, user needs to provide \textbf{\textit{k}}-mesh size to the \textit{kgrid} variable in the input file. For instance, if \textbf{\textit{k}}-mesh size of 2$\times$2$\times$2 is required, one should assign the string \textquotedblleft 2 2 2\textquotedblright \hspace*{0.05in} to \textit{kgrid} variable. In addition to this, if shifted \textbf{\textit{k}}-mesh is needed, one must allocate 1 to \textit{shift\_in\_k} variable, else it must be assigned 0. One must note that the shifting of \textbf{\textit{k}}-points is done exactly as implemented in the WIEN2k package\cite{blaha2020wien2k}. Based on these inputs and the value assigned to \textit{struct\_num} variable, \textit{PY-BerryAHC} internally generates \textit{case.klist} file. It is suggested to use the shifted \textbf{\textit{k}}-mesh for AHC calculations using the \textit{PY-BerryAHC} code. Moving further, it is generally observed in WIEN2k calculations that while calculating the eigenvalues (specifically, while running \textit{x lapw1}) the amount of memory occupied on RAM increases with the increase in number of \textbf{\textit{k}}-points in the \textit{case.klist} file. Thus, beyond a certain size of \textbf{\textit{k}}-mesh, RAM gets fully occupied and the calculations get aborted. The size of \textbf{\textit{k}}-mesh beyond which it happens varies from system to system depending on the size of RAM. To avoid this situation and for smooth running of $\boldsymbol\Omega$ calculations, the input file is provided with a variable named $num\_kp\_at\_once$. The original \textit{case.klist} file is renamed as \textit{case.klist1}. Now, based on the value assigned to $num\_kp\_at\_once$, that number of \textbf{\textit{k}}-points will be extracted from the \textit{case.klist1} file, and a new \textit{case.klist} file will be created. Using this file, eigenvalues and momentum-matrices are calculated via WIEN2k following which $\boldsymbol\Omega$ is calculated and stored. Having this done, again a new \textit{case.klist} file is created with next $num\_kp\_at\_once$ number of \textbf{\textit{k}}-points from \textit{case.klist1} file and the process is repeated until all the \textbf{\textit{k}}-points are processed. With this, the discussion of all the input variables is complete. Now, a detailed workflow of \textit{PY-BerryAHC} is discussed below.

The flowchart of the processing of \textit{PY-BerryAHC} code is shown in figure \ref{workflow}. At the very first, \textit{PY-BerryAHC} generates the \textit{case.klist} file based on the \textbf{\textit{k}}-mesh size assigned to \textit{\textbf{k}}-grid variable. Moving next, if calculation is to be done in serial mode ($NUM\_THREADS$=1), the code does not further divide the \textit{case.klist} file. However, if computation is to be done in parallel mode ($NUM\_THREADS>$1), it further divides the file. Lets say if $NUM\_THREADS$=4. Then the code will first divide the \textbf{\textit{k}}-points of \textit{case.klist} file into 4 smaller files with the name- \textit{case.klist1}, \textit{case.klist2}, ..., \textit{case.klist4}. Each of these files will contain equal number of \textit{\textbf{k}}-points. However, if some \textit{\textbf{k}}-points remain extra after this equal distribution, they are adjusted in the last \textit{case.klistX} file. After this, \textit{PY-BerryAHC} creates a \textit{berry\_calc\_dir} directory. Within this directory, it will create $NUM\_THREADS$ number of sub directories with names- \textit{process\_xxxx..}, where \textit{xxxx..} are randomly assigned names by the system. This helps the system to keep track of the results obtained from each of the jobs running on different threads in parallel calculations. These process IDs are also written in a file named $jobs\_location.txt$. Within each \textit{process\_xxxx..} folder, the case folder along with the $PY-BerryAHC.input$ and $case.inop$ files are copied. Then the $BerryAt\_kpt.py$ module is called which calculates the eigenvalues and momentum-matrices at each \textbf{\textit{k}}-points. After this, $BerryAt\_kpt.py$ will call another module named $master\_code.py$ which will read the eigen energy (from \textit{case.energyso} file) and momentum-matrices (from \textit{case.pmat} file) and computes the bands-resolved values of $\Omega_{x}$, $\Omega_{y}$ \& $\Omega_{z}$. The results will be stored in $case.berry\_curv\_full$ file. This process will be repeated for all the \textbf{\textit{k}}-points in \textit{case.klist} file at each threads. Having done this, everything is set for the post-process (calculation of $\sigma_{\mu \nu}^{\mathrm{AHC}}$).

\textbf{\underline{Post-process:}} The post-process is carried out by calling the command: \textit{python3 \$BerryAHC/post\_process.py}. In the post-process, \textit{PY-BerryAHC} performs the calculations of $f_n(\textbf{\textit{k}}) \Omega_\xi^n(\textbf{\textit{k}})$ corresponding to all the bands at each \textbf{\textit{k}}-points. In a single run, it computes $f_n(\textbf{\textit{k}}) \Omega_\xi^n(\textbf{\textit{k}})$ at large number of $\mu$ values in a given chemical potential window. Also, for each of these $\mu$ values, it computes AHC in a range of temperature values at the same time. As already mentioned, the chemical potential window is decided by the values assigned to variables- \textit{chemical\_potential}, \textit{emin}, \textit{emax} and \textit{estep} while the temperature window is defined by the values allocated to the variables- \textit{Tmin}, \textit{Tmax} and \textit{Tstep}.  Finally, it computes $\sigma_{\mu \nu}^{\mathrm{AHC}}$ (for $\mu \nu$= {$xz$, $yz$ \& $xy$}) at each value of $\mu$ by summing the computed values of $f_n(\textbf{\textit{k}}) \Omega_\xi^n(\textbf{\textit{k}})$ at each \textbf{\textit{k}}-points sampled across the BZ. The result is stored in \textit{sigma.txt} file. One must note that while running the \textit{PY-BerryAHC} module (\textit{python3 \$BerryAHC/PY-BerryAHC}), the post-process is automatically called at the end of the $\boldsymbol\Omega$ calculations. However, once the $\boldsymbol\Omega$ file is obtained, one can run the post-process independently to calculate the $\sigma_{\mu \nu}^{\mathrm{AHC}}$ at different values of $T$ or $\mu$. The post-process is required to be run in serial or parallel mode depending on the mode in which $\boldsymbol\Omega$ calculations have been carried out. The value of $NUM\_THREADS$ must also be kept same. Before ending this section, it is important to note that the total time taken for $\boldsymbol\Omega$ calculation is mentioned in \textit{final\_time.dat file}. In addition to this, the time taken in the post-process of AHC computation in mentioned at the end of \textit{sigma.txt} file. This is useful for having track of time taken for the calculations.

\begin{figure*}[t]
    \centering
    \begin{subfigure}[b]{0.45\textwidth}
        \includegraphics[width=\textwidth]{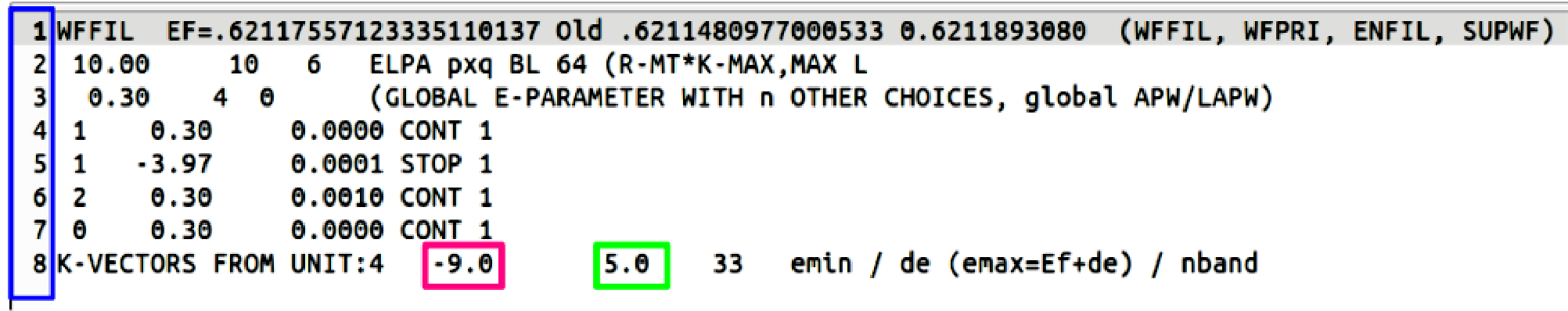}
        \caption{\textit{case.in1} file.}
        \label{fig:sub1}
    \end{subfigure}
    \hfill
    \begin{subfigure}[b]{0.45\textwidth}
        \includegraphics[width=\textwidth]{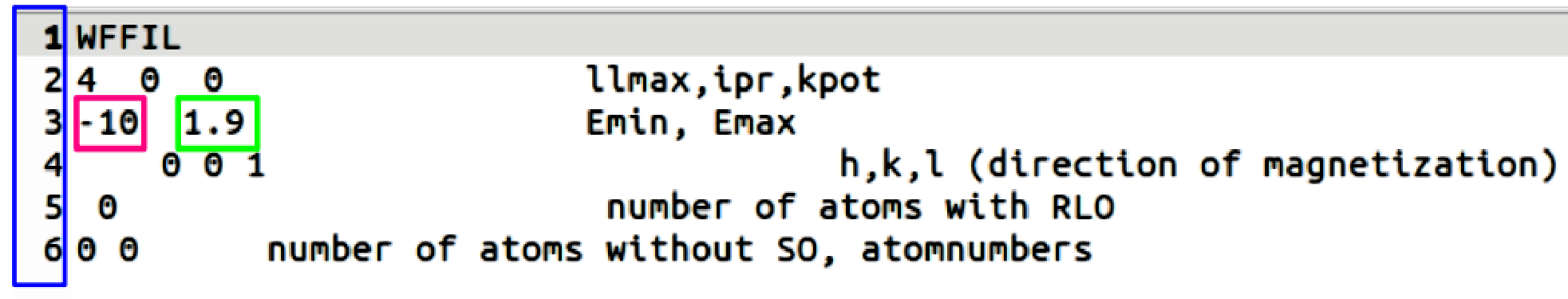}
        \caption{\textit{case.inso} file.}
        \label{fig:sub2}
    \end{subfigure}
    \begin{subfigure}[b]{0.45\textwidth}
        \includegraphics[width=\textwidth]{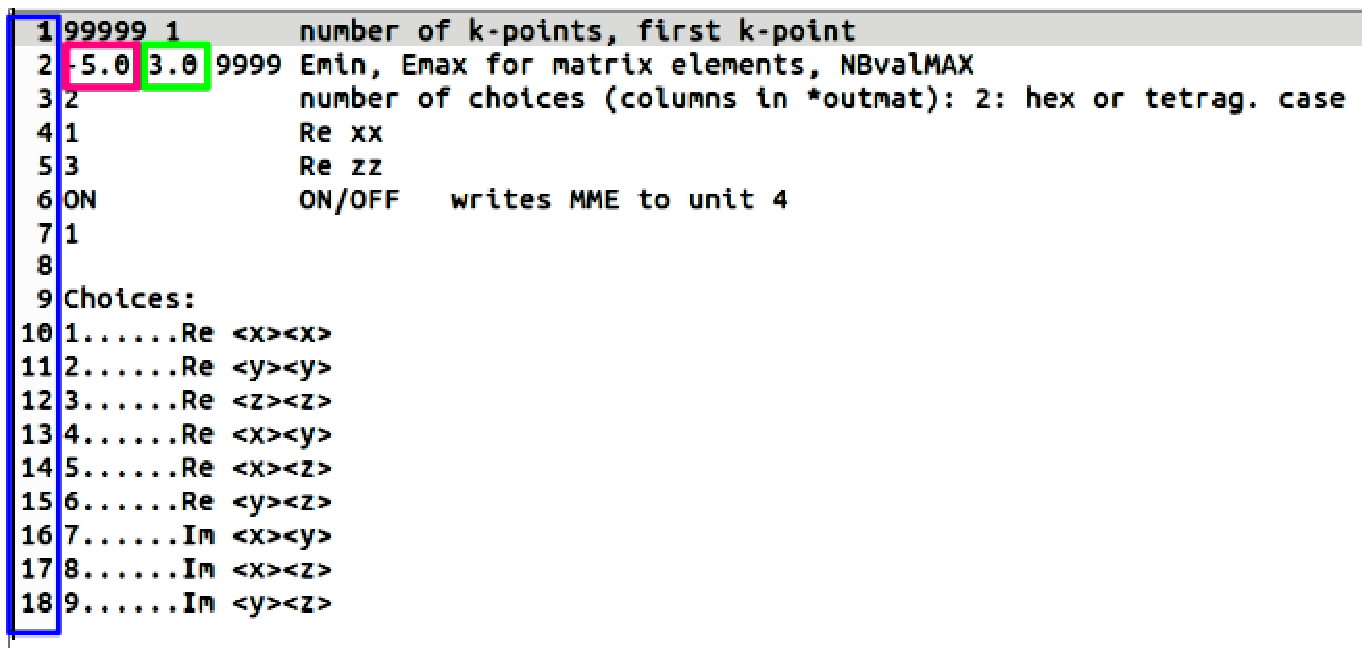}
        \caption{\textit{case.inop} file.}
        \label{fig:sub2}
    \end{subfigure}
    \caption{WIEN2k input files for eigenvalues and momentum-matrices calculations.}
    \label{figFiles}
\end{figure*}

\subsection{Technical details}
All the modules of \textit{PY-BerryAHC} code is written in Python 3. Hence, it may not run with the lower versions of python. Various modules of python like- \textit{numpy}, \textit{math}, \textit{time}, \textit{multiprocessing}, \textit{uuid}, \textit{struct}, etc have been explicitly used in it. Moving next, the calculation of $\boldsymbol\Omega$ (and AHC in the post-process) is made parallel over the \textbf{\textit{k}}-points using the \textit{Pool} class of \textit{multiprocessing} module of Python 3. Furthermore, calculation of $\boldsymbol\Omega$ \& AHC is done using the \textit{first-principle} approach. Current version of the code is interfaced with the WIEN2k package. However, it can be easily interfaced with the other DFT packages like- ELK\cite{Elk}, ABINIT\cite{gonze2016recent}, \textit{etc}. Also, the code has been designed in its most general form without considering any specific class of material or model. This ensures its wide range of applicability and practical use.\\\\

\textbf{\underline{Strategy for efficient use of the code}:} A close inspection of equations \ref{eq2} \& \ref{eq31} suggest that for a given value of $T$, a pair of bands will not contribute to AHC if at each \textbf{\textit{k}}-points of the BZ, the occupancy of one band is exactly equal to the other band. The pair of bands will only have non-zero contribution to total AHC at the given value of $T$, if in some regions of the BZ, the occupancy of one of the band differs from the other. Next thing that these equations suggest is that if a pair of bands have large energy difference, then their contributions to AHC will be negligible and thus can be safely ignored in computations. This discussion suggests that calculations of all the bands at a given \textbf{\textit{k}}-points, which generally consumes higher computational resources is not really required. It is therefore suggested that for upto temperature of 300 K, AHC at a given chemical potential ($\mu$) can be computed by taking the energy interval of not more than $\pm$1.5 eV with respect to $\mu$. The procedure to achieve this in WIEN2k calculations is discussed further.

\begin{table*}
\caption{\label{tab:table1}%
\normalsize{The various input details and the \textit{k}-mesh size used to calculate the ground state energy of the materials using WIEN2k package. The \textit{\textbf{k}}-meshes are taken in the irreducible part of the Brillouin zone (IBZ).
}}
\begin{ruledtabular}
\begin{tabular}{ccccc}
\textrm{$\textbf{case}$}&
\textrm{$\textbf{space-group}$}&
\textrm{$\textbf{lattice parameters}$}&
\textrm{$\textbf{Wyckoff Positions}$}&
\textrm{\textbf{\textit{k}-mesh (in IBZ)}}\\
\colrule\\
    Fe\cite{Fe-lp}   &$Im-3m$   & a=b=c=2.87 \AA      & Fe = (0.00, 0.00, 0.00)             &  10$\times$10$\times$10 \\
                     &          & $\alpha$=$\beta$=$\gamma$=90       &           &                       \\\\
Fe$_3$Ge\cite{drijver1976magnetic,kanematsu1963magnetic,cao2009large} &$Fm\bar{3}m$ & a=b=5.76 \AA       & Fe (I) = (0.75,0.25,0.75)        & 10$\times$10$\times$10  \\
         &           & $\alpha$=$\beta$=$\gamma$ 90 & Fe (II) = (0.00, 0.00, 0.50)        &                         \\
         &           &                                    & Ge = (0.00, 0.00, 0.00)              &                         \\\\
Co$_2$FeAl\cite{shukla2022atomic} &$Fm\bar{3}m$ & a=b=c=5.70 \AA              & Co = (0.25, 0.25, 0.25)        & 10$\times$10$\times$10  \\
         &           & $\alpha$=$\beta$=$\gamma$= 90     & Fe = (0.50, 0.50, 0.50)        &                         \\
         &           &                                   & Al = (0.00, 0.00, 0.00)              &                         \\\\
\end{tabular}
\end{ruledtabular}
\label{tab3}
\end{table*}

\begin{table*}
\caption{\label{tab:table1}%
\normalsize{The details of the values assigned to the input parameters for the \textit{PY-BerryAHC} code. The \textbf{\textit{k}}-grid is sampled across the full Brillouin zone (FBZ)..
}}
\begin{ruledtabular}
\begin{tabular}{ccccccc}
\textrm{$\textbf{\textit{case}}$}&
\textrm{$\textbf{\textit{struct\_num}}$}&
\textrm{$\textbf{\textit{Temperature}}$}&
\textrm{$\textbf{\textit{chemical\_potential}}$}&
\textrm{$\textbf{\textit{spin\_pol}}$}&
\textrm{$\textbf{\textit{kgrid}}$}&
\textrm{$\textbf{\textit{shift\_in\_k}}$}\\
\textrm{$\textbf{ }$}&
\textrm{$\textbf{ }$}&
\textrm{$\textbf{(Kelvin)}$}&
\textrm{$\textbf{(Rydberg)}$}&
\textrm{$\textbf{ }$}&
\textrm{$\textbf{ }$}&
\textrm{$\textbf{ }$}\\
\colrule
    Fe     & 3 & 300 & 0.6211893080 & 1 & 400$\times$400$\times$400 & 1\\
Fe$_3$Ge   & 2 & 300 & 0.6681367548 & 1 & 400$\times$400$\times$400 & 1\\
Co$_2$FeAl & 2 & 2   & 0.6276267143 & 1 & 400$\times$400$\times$400 & 1\\
\end{tabular}
\end{ruledtabular}
\label{tab31}
\end{table*}

For calculating AHC, one needs the bands\textquoteright \hspace*{0.02in} energies and their momentum-matrices at each \textbf{\textit{k}}-points. Thus, for better efficiency of the \textit{PY-BerryAHC} code, one needs to control the computations of these quantities within a given energy window (close to Fermi energy for general purpose). For calculating the eigenvalues of only the bands falling within a given energy window, one needs to specify the lower energy limit (in absolute unit) and higher energy limit (with respect to Fermi energy) in the \textit{case.in1} file. Figure \ref{figFiles} (a) shows a sample \textit{case.in1} file. The line numbers in the file is marked in the blue box. The position of lower energy limit (in absolute unit in Rydberg) \& higher energy limit (in Rydberg and scaled with respect to Fermi energy) is marked inside the red \& green boxes, respectively. Moving further, one also needs to limit the calculation of SOC corrected eigenvalues within the required energy window. This can be done by specifying the upper and lower limits of energy window in absolute unit (in Rydberg) in the \textit{case.inso} file. The sample \textit{case.inso} file is shown in figure \ref{figFiles} (b) with upper \& lower limit of energy window marked in red \& blue boxes, respectively. Apart from the eigenvalues, one should also limit the energy window for the calculations of momentum-matrices. This can be done by specifying the upper and lower energy limits in absolute units (in Rydberg) in \textit{case.inop} file as shown in figure \ref{figFiles} (c). Again, the position of upper \& lower energy limits is marked in red \& green boxes, respectively. It is also important to highlight here that \textit{PY-BerryAHC} reads the momentum-matrices from the \textit{case.pmat} file generated by the OPTIC module of WIEN2k package. \textit{case.inop} file is the input file for the OPTIC module. However, the default version of this file will not produce the \textit{case.pmat} file. For generating it, user is required to explicitly write \textquoteleft \textit{ON}\textquoteright\hspace*{0.02in}  in line 6 and \textquoteleft \textit{1}\textquoteright\hspace*{0.02in}  in line 7 of the \textit{case.inop} file. This is also shown in figure \ref{figFiles} (c). Following these strategies, \textit{PY-BerryAHC} is expected to efficiently calculate AHC for any given material.

\section{Test Cases}
For validating the code, it has been tested over three ferromagnetic materials: Fe\cite{Fe-lp}, Fe$_3$Ge\cite{drijver1976magnetic,kanematsu1963magnetic,cao2009large} and Co$_2$FeAl\cite{shukla2022atomic}. Fe is a body-centered cubic crystal while the other two materials crystallize in face-centered cubic crystal structure. Having intrinsic magnetism, these materials are expected to exhibit AHE. In this regard, several theoretical calculations and experimental measurements have been previously carried out\cite{danan1968new,dheer1967galvanomagnetic,wang2006ab,li2023anomalous,shukla2022atomic,huang2015anomalous}. These studies have verified the AHE in the materials and reported the values of AHC at various temperatures. Thus, it is convincing to use them for benchmarking the \textit{PY-BerryAHC} code.

\begin{figure*}[t]
    \centering
    \begin{subfigure}[b]{0.30\textwidth}
        \includegraphics[width=\textwidth]{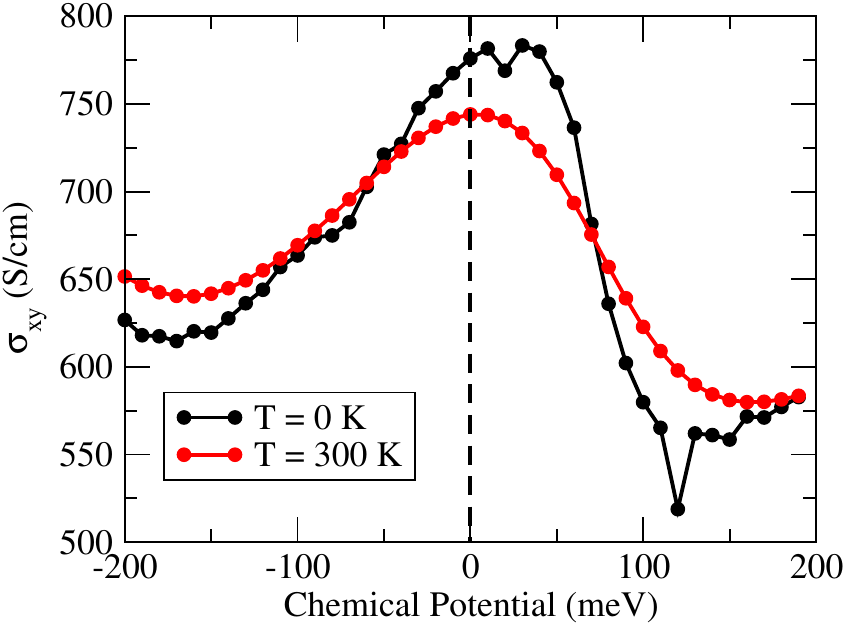}
        \caption{Fe}
        \label{sigmaFe}
    \end{subfigure}
    \begin{subfigure}[b]{0.30\textwidth}
        \includegraphics[width=\textwidth]{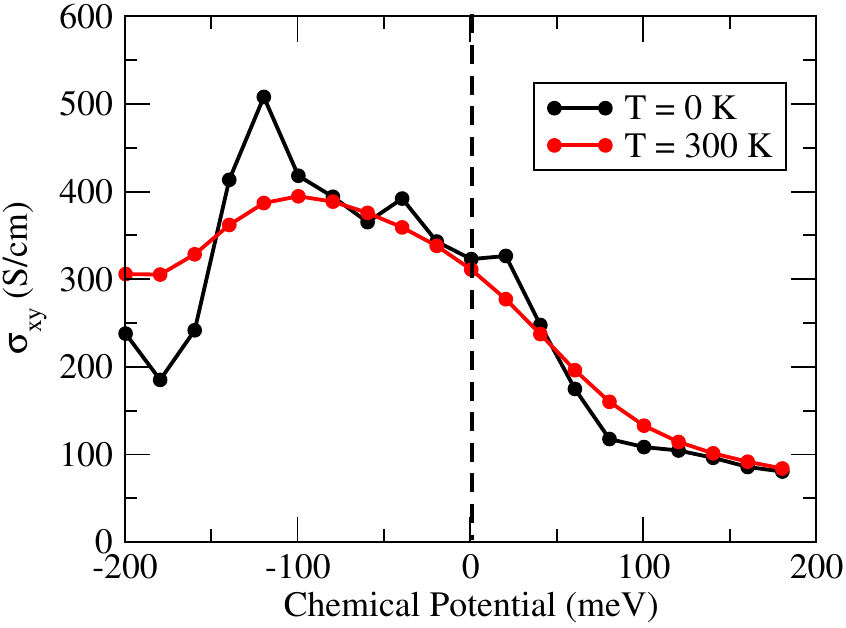}
        \caption{Fe$_3$Ge}
        \label{sigmafe3ge}
    \end{subfigure}
    \begin{subfigure}[b]{0.30\textwidth}
        \includegraphics[width=\textwidth]{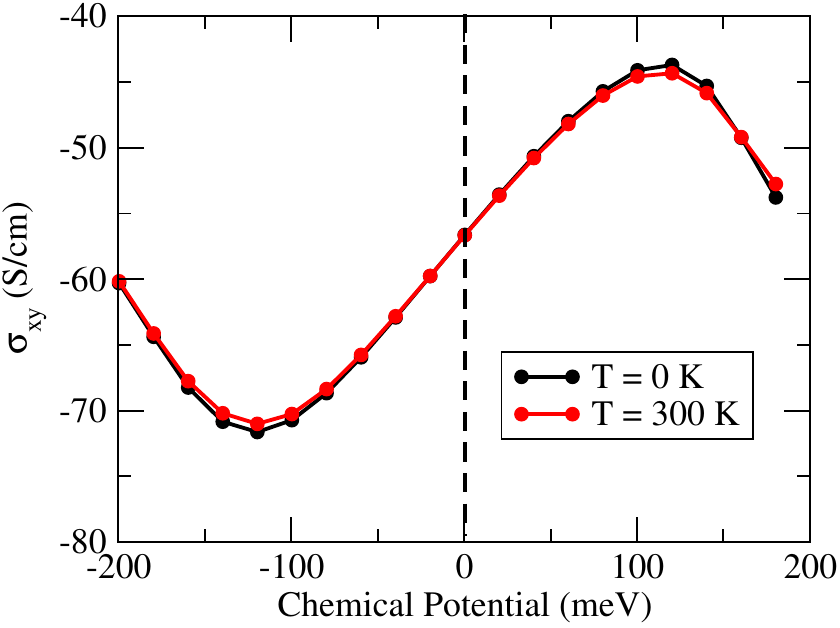}
        \caption{Co$_2$FeAl}
        \label{figco2feal}
    \end{subfigure}
    \caption{The black (red) color curve represent the $\sigma_{xy}$ vs chemical potential for respective materials at 0 K (300 K). The chemical potential is scaled with respect to Fermi energy.}
    \label{figFiles2}
\end{figure*}
 
\subsection{Computational Details}
The self-consistent ground state energy calculations for all the materials are performed using the WIEN2k package. SOC has been included in the ground state energy calculations. Moreover, all the materials are magnetic in nature. So, spin-polarization effect is also included in these computations with magnetization taken in the \textit{z} direction. Furthermore, Perdew-Burke-Ernzerhof (PBE)\cite{perdew1996generalized}, which is based on generalized gradient approximation (GGA) is used as exchange-correlation functional in these computations. Space group, lattice parameters, Wyckoff positions of atoms and \textit{\textbf{k}}-mesh size used for ground-state computations of these materials are mentioned in table \ref{tab3}. In all the calculations, the \textit{\textbf{k}}-mesh is sampled in the irreducible parts of the BZ. Furthermore, the energy convergence limit for self-consistent ground state energy calculation is set to 10$^{-4}$ Rydberg per unit cell for all the calculations.

For the AHC computations, \textbf{\textit{k}}-mesh size and the values assigned to other input parameters of \textit{PY-BerryAHC} code are mentioned in table \ref{tab31}. For all the calculations, the value of $\sigma_{xy}$ were computed for the chemical potential window ranging from -200 to 200 meV with respect to Fermi level. This chemical potential window was divided into 20 equally spaced parts. Thus, values of \textit{emin}, \textit{emax} and \textit{estep} were set to -200, 200 \& 20. In addition to this, all the computations were performed on a system with 64 GB RAM size. In this configuration, the values assigned to the parameters \textit{NUM\_THREADS} \& \textit{num\_kp\_at\_once} are 28 \& 5000, respectively.

\begin{table}
\caption{\label{tab:table1}%
\normalsize{Convergence of $\sigma_{xy}$ at 300 K with respect to the size of \textbf{\textit{k}}-mesh in Fe. The chemical potential is set at Fermi level.
}}
\begin{ruledtabular}
\begin{tabular}{cc}
\textrm{\textbf{\textit{k}-mesh}}&
\textrm{\textbf{$\sigma_{xy}$}}\\
\textrm{\textbf{}}&
\textrm{\textbf{$S/cm$}}\\
\colrule
 50$\times$50$\times$50              & 831.96 \\
 100$\times$100$\times$100           & 756.88 \\
 150$\times$150$\times$150           & 739.12 \\
 200$\times$200$\times$200           & 744.95 \\
 250$\times$250$\times$250           & 740.41 \\
 300$\times$300$\times$300           & 743.75 \\
 350$\times$350$\times$350           & 743.46   \\
 400$\times$400$\times$400           & 743.49     \\
    
\end{tabular}
\end{ruledtabular}
\label{tab4}
\end{table}

\begin{table}
\caption{\label{tab:table1}%
\normalsize{Magnitude of $\sigma_{xy}$ vs Temperature. The chemical potential is set at Fermi level.
}}
\begin{ruledtabular}
\begin{tabular}{cccc}
\textrm{\textbf{Temperature}}&
\textrm{\textbf{}}&
\textrm{\textbf{$\sigma_{xy}$}}&\\
\colrule
\textrm{\textbf{}}&
\textrm{\textbf{Fe}}&
\textrm{\textbf{Fe$_3$Ge}}&
\textrm{\textbf{Co$_2$FeAl}}\\
\textrm{\textbf{(Kelvin)}}&
\textrm{\textbf{($S/cm$)}}&
\textrm{\textbf{($S/cm$)}}&
\textrm{\textbf{($S/cm$)}}\\
\colrule
 0            & 775.94  & 322.90 & 56.15 \\
 25           & 774.61  & 323.10 & 56.29 \\
 50           & 773.33  & 324.44 & 56.48 \\
 75           & 772.17  & 326.08 & 56.49  \\
 100          & 770.78  & 326.99 & 56.49   \\
 125          & 768.97  & 326.80 & 56.49  \\
 150          & 766.62  & 325.63 & 56.51    \\
 175          & 763.72  & 323.74 & 56.52   \\
 200          & 760.35  & 321.42 & 56.55  \\
 225          & 756.59  & 318.87 & 56.57  \\
 250          & 752.55  & 316.24 & 56.60  \\
 275          & 748.32  & 313.60 & 56.63 \\
 300          & 744.00  & 311.00 & 56.66   \\
    
\end{tabular}
\end{ruledtabular}
\label{tabb4}
\end{table}

\subsection{Results and discussion}
To estimate the size of \textit{\textbf{k}}-mesh required to obtain the converged AHC value using the \textit{PY-BerryAHC} code, the convergence of $\sigma_{xy}$ with respect to the size of \textbf{\textit{k}}-mesh is carried out for Fe. In these calculations, chemical potential and temperature values were set at the Fermi level and 300 K, respectively. Obtained results are depicted in table \ref{tab4}. It is seen from the table that beyond the \textbf{\textit{k}}-mesh size of 350$\times$350$\times$350, the value of $\sigma_{xy}$ is almost constant with the increase in size of the \textit{\textbf{k}}-mesh. For maintaining the accuracy of the calculated results, the \textit{\textbf{k}}-mesh size of 400$\times$400$\times$400 is taken for all further calculations of AHC for validation of the code. Moving further, computation of the temperature dependent value of $\sigma_{xy}$ at the Fermi level has been also carried out for Fe. The temperature range chosen is from 0-300 K. The obtained results are mentioned in table \ref{tabb4}. It is seen from the table that value of $\sigma_{xy}$ decreases with the rise in temperature. In addition to this, the calculations have been also carried out to study the variation of $\sigma_{xy}$ with the change in chemical potential ($\mu$). The values have been calculated for a range of $\mu$ values ranging from -200 meV to 200 meV with respect to Fermi energy. Furthermore, this calculation has been performed at 0 \& 300 K. Obtained results are shown in figure \ref{sigmaFe}. It is observed that, corresponding to both the temperature values, as $\mu$ value is raised above (or lowered below) the Fermi energy, magnitude of $\sigma_{xy}$ decreases. At the Fermi energy, magnitude of $\sigma_{xy}$ is obtained to be $\sim$775 ($\sim$744) $S/cm$ at 0 (300) K. Yao \textit{et. al.} have previously calculated the value of $\sigma_{xy}$ using the \textit{first-principle} approach combined with the adaptive mesh refinement method\cite{yao2004first}. The value of $\sigma_{xy}$ reported in their work is 751 (734) $S/cm$ at 0 (300) K. Apart from this, Dheer \textit{et. al.} performed experiments on iron whiskers and extracted the value of $\sigma_{xy}$ at room temperature\cite{dheer1967galvanomagnetic}. The obtained value as reported in their work was 1032 $S/cm$. The value of $\sigma_{xy}$ calculated from \textit{PY-BerryAHC} code is found to be in good agreement with these reported data.

Moving to the case of Fe$_3$Ge, the temperature dependent value of $\sigma_{xy}$ for $\mu$ corresponding to Fermi level is computed using \textit{PY-BerryAHC} code. Obtained results are depicted in table \ref{tabb4}. It is seen from the table that with the rise in temperature from 0 to 100 K, magnitude of $\sigma_{xy}$ slowly increases from $\sim$322 $S/cm$ to $\sim$326 $S/cm$. With further increase in temperature, magnitude of $\sigma_{xy}$ falls down slowly. At 300 K, the computed value of $\sigma_{xy}$ is found to be 311 $S/cm$. Li \textit{et. al.} have experimentally explored the temperature dependence of AHC in polycrystalline Fe$_3$Ge\cite{li2023anomalous}. At room temperature, the measured value of AHC as reported in their work is 106 \textit{S/cm}. In addition to this, they also explored AHC of Fe$_3$Ge using theoretical approach based on wannierization techniques. The value of AHC was calculated to be 227 \textit{S/cm} at 300 Kelvin. It was explained that since the sample was polycrystalline, the experimental value of AHC was found to be smaller than theoretically calculated one. Keeping the polycrystalline aspect in mind and the wannierization techniques being used in their theoretical computations, the obtained value of $\sigma_{xy}$ from \textit{PY-BerryAHC} code is in good match with their reported data. In addition to this, the effect of change in $\mu$ value on $\sigma_{xy}$ have been also explored for the material at 0 K \& 300 K using the \textit{PY-BerryAHC} code. Obtained results are shown in figure \ref{sigmafe3ge}. It is seen that, within the given range of $\mu$ and at 0 K \& 300 K, on raising the value of $\mu$ from -200 meV (with respect to Fermi energy), magnitude of $\sigma_{xy}$ first increase till $\mu$ value reaches close to -100 meV (with respect to Fermi energy). However, with further rise in the value of $\mu$, magnitude of $\sigma_{xy}$ falls down.

\begin{figure}[t]
    \centering
    \includegraphics[width=0.45\textwidth,height=5.0cm]{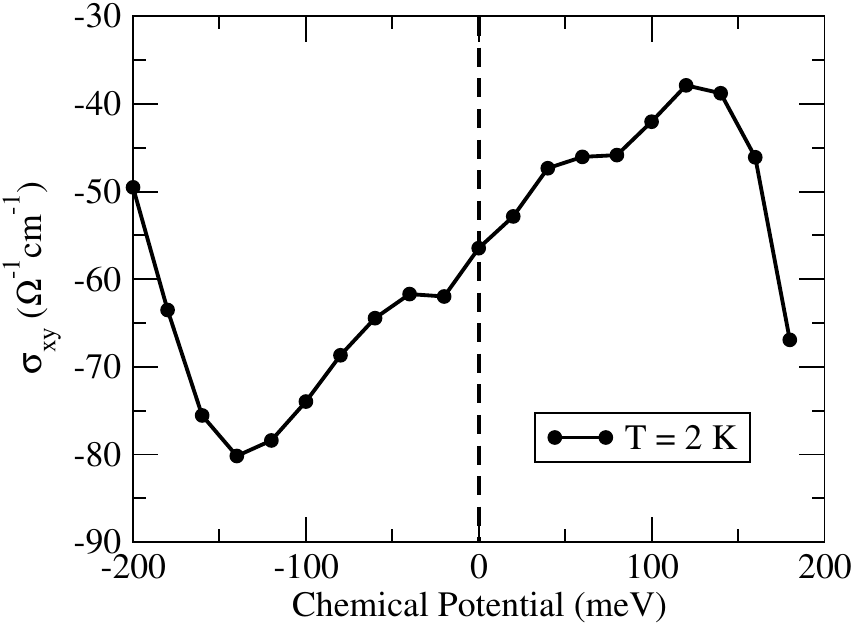}
    \caption{$\sigma_{xy}$ vs chemical potential for Co$_2$FeAl at 2 K. The chemical potential is scaled with respect to Fermi energy. }
    \label{sigmaCo2FeAl}
\end{figure}

In case of Co$_2$FeAl, the temperature-dependent variation of magnitude of $\sigma_{xy}$ is studied using the \textit{PY-BerryAHC} code. The obtained results are mentioned in table \ref{tabb4}. It is found that magnitude of $\sigma_{xy}$ is nearly constant with the change in temperature within 0-300 K. In addition to this, the effect of change in the value of $\mu$ on $\sigma_{xy}$ of Co$_2$FeAl is studied for temperature values of 0 K \& 300 K. Obtained results are depicted in figure \ref{figco2feal}. It is observed that, within the given $\mu$-window, $\sigma_{xy}$ at 0 K \& 300 K are very close to each other. Moreover, $\sigma_{xy}$ show slow variation with the change in $\mu$ within the given $\mu$-window. In the literature survey, we came across a work in which $\sigma_{xy}$ is calculated at 2 K (42 $S/cm$\cite{shukla2022atomic}). Thus, calculation of $\sigma_{xy}$ is carried out at 2 K using \textit{PY-BerryAHC} code. Obtained result is shown in figure \ref{sigmaCo2FeAl}. At the Fermi level, the magnitude of $\sigma_{xy}$ is found to be $\sim$56 $S/cm$ at 2 K. The features and magnitudes of the result is in good agreement with the previously reported theoretical result (39 $\Omega^{-1}cm^{-1}$\cite{huang2015anomalous}) at 2 K. The experimental reported value of $\sigma_{xy}$ is 155 $\Omega^{-1}cm^{-1}$\cite{shukla2022atomic}, which was found to be nearly constant with the variation in temperature. Our results also matches to the trend of this experimental observation.

Before concluding this section, it is important to mention that the magnitude of AHC computed using \textit{PY-BerryAHC} may sometime be found to largely deviate from the experimental results, as in the case of Co$_2$FeAl. The possible reason behind this may be the impurity or defect in the sample under experimental investigation. Such impurity or defect in material gives rise to various kind of scattering mechanism thereby affecting the total AHC of the compound.

\begin{figure*}[t]
    \centering
    \includegraphics[width=0.85\textwidth,height=5.0cm]{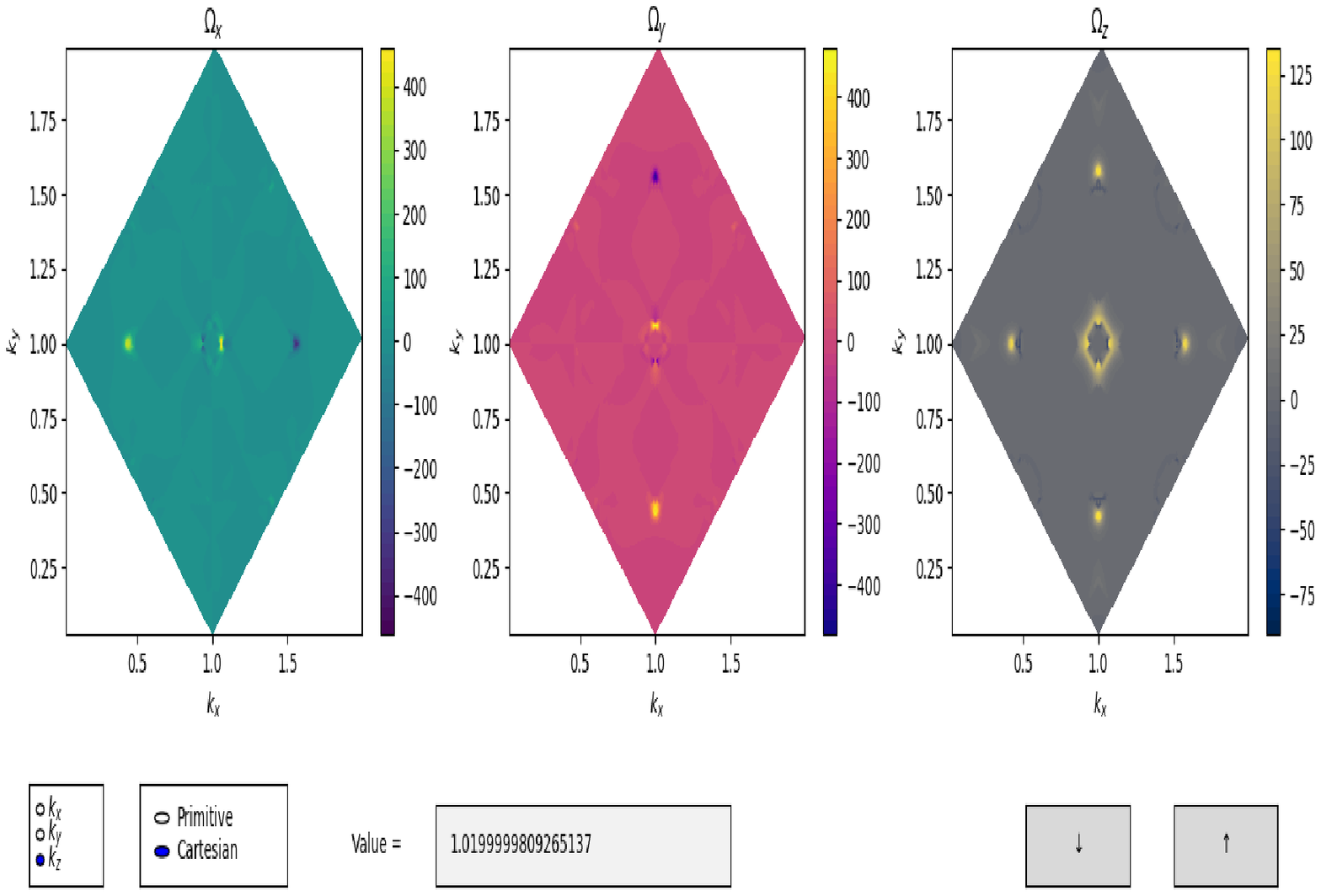}
    \caption{The interactive window produced by \textit{berry\_plot.py} module for Fe (at 0 K and chemical potential set to Fermi level). It displays the $\Omega_x(\textbf{\textit{k}})$, $\Omega_y(\textbf{\textit{k}})$ \& $\Omega_z(\textbf{\textit{k}})$  at $k_z$= $\sim$1.02 in cartesian coordinate system.}
    \label{berryPlotFe}
\end{figure*}

\section{The \textit{berry\_plot.py} module}
It is seen from equation \ref{eq31} that the term $\sum_n f_n(\textbf{\textit{k}}) \Omega_\xi^n(\textbf{\textit{k}})$ palys a vital role in deciding the extent to which a \textbf{\textit{k}}-point will contribute to total AHC (at a given value of $T$ and $\mu$) of any material under study. In this regard, it is important to mention here that this contribution may get canceled out by the contribution from a term with equal magnitude but opposite sign at some other \textbf{\textit{k}}-point of the BZ. Thus, study of variation of this term across the BZ may give some deeper insight regarding the origin of AHC in any given material. Such analysis is specially important in studying WSMs where AHC is claimed to get enhanced due to the states close to Weyl nodes\cite{shen2022intrinsically,zhou2020enhanced}. For such analysis, \textit{PY-BerryAHC} is equipped with a module named \textit{berry\_plot.py}. This module must be used as post-processing tools after the calculation of $\boldsymbol\Omega$. Also, for just the visual analysis of the variation of $\boldsymbol\Omega$, the computation can be done on smaller size of \textbf{\textit{k}}-mesh (50$\times$50$\times$50 or 100$\times$100$\times$100) as compared to AHC computation. Various functions of this module and steps to use it is discussed further.

The module calculates $\Omega_\xi(\textbf{\textit{k}})$=$\sum_n f_n(\textbf{\textit{k}}) \Omega_\xi^n(\textbf{\textit{k}})$, where $\xi$=$x$, $y$ or $z$. The temperature and chemical potential value corresponding to which the calculation is to be done must be assigned to variables \textit{berry\_plot\_at\_T} and \textit{chemical\_potential}, respectively. After this, one must run the command- \textit{python3 \$BerryAHC/berry\_plot.py} to perform the calculations. Having done the computations, it generates an interactive window displaying $\Omega_x(\textbf{\textit{k}})$, $\Omega_y(\textbf{\textit{k}})$ \& $\Omega_z(\textbf{\textit{k}})$ in $k_x$=$C$ or $k_y$=$C$ or $k_z$=$C$ plane. The screenshot of the window obtained on performing the calculations on Fe at 0 Kelvin and chemical potential set at Fermi level, is shown in figure \ref{berryPlotFe}. As can be seen in the figure, the window provides various interactive options for further analysis of $\Omega_\xi(\textbf{\textit{k}})$. For instance, it provides the option to switch between $k_x$=$C$, $k_y$=$C$, and $k_z$=$C$ planes. In addition to this, it also provides an option to change the value of $C$ using the arrow key as shown in the figure. This helps in better visualization of $\Omega_\xi(\textbf{\textit{k}})$. The value of $C$ at which $\Omega_\xi(\textbf{\textit{k}})$ term is shown at any instance is displayed in the text box with label \textit{Value}. Apart from this, many a times one needs to switch from primitive coordinate system to cartesian coordinate system or vice-versa. For instance, if one has the coordinates of Weyl nodes in either primitive coordinate system or cartesian coordinate system and need to verify how $\boldsymbol\Omega$ behaves around the Weyl node. Keeping these aspects in mind, the window is also provided with an option to switch from primitive to cartesian coordinate system or vice-versa. One must note that this conversion is only implemented for the structures mentioned in table \ref{tab2}. The features provided by \textit{berry\_plot.py} module is expected to be very useful for $\boldsymbol\Omega$ \& AHC analysis in any material and to have a better understanding of the origin of AHC in it. Thus, the module greatly enhance the capability of \textit{PY-BerryAHC} package in exploring topological materials.

\section{Conclusion}
In the present work, a Python 3 code named \textit{PY-BerryAHC} has been designed to calculate anomalous Hall conductivity (AHC) of any material using the \textit{first-principle} approach. The code employs Kubo-like formula to calculate the Berry curvature ($\boldsymbol\Omega$) by using the eigenvalues and momentum-matrices obtained from WINE2k calculations. Parallel computing of $\boldsymbol\Omega$ \& AHC along with the efficient handling of band-resolved $\boldsymbol\Omega$ in binary format is found to greatly enhance the efficiency and usability of the code. For benchmarking of the code, it has been tested on some well-studied ferromagnetic materials exhibiting AHC. These include- Fe, Fe$_3$Ge \& Co$_2$FeAl. The values of AHC of these materials, calculated from AHC code, are found to be in good match with those reported in literature. For instance, for the case of Fe, the magnitude of $\sigma_{xy}$ when chemical potential ($\mu$) is set at the Fermi energy is obtained to be $\sim$775 ($\sim$744) $S/cm$ at 0 (300) K. These values are found to be in close agreement to previously reported theoretical data (751 (734) $S/cm$ at 0 (300) K\cite{yao2004first}) and experimental results (1032 $S/cm$ at 300 K\cite{dheer1967galvanomagnetic}). Moving further, in case of Fe$_3$Ge, the calculated value of $\sigma_{xy}$ is found to be 311 $S/cm$ at the room temperature. This result is also found to be in very good match with the previously reported experimental data (106 $S/cm$) and wannierization-based theoretical computation (227 $S/cm$) at 300 K\cite{li2023anomalous}. Nextly, for Co$_2$FeAl, the magnitude of computed value of $\sigma_{xy}$ at 2 K when the chemical potential is set at the Fermi level is found to be $\sim$56 $S/cm$. This is also in good agreement with the previously reported theoretical result (42 $S/cm$\cite{shukla2022atomic}, 39 $S/cm$\cite{huang2015anomalous}) and the experimental data (155 $S/cm$\cite{shukla2022atomic}) at 2 K. These results validate the accuracy, efficiency and reliability of \textit{PY-BerryAHC} code in computing AHC of any given material. Also, the visualization tool (\textit{berry\_plot.py}) for $\sum_n f_n(\textbf{\textit{k}}) \Omega_\xi^n(\textbf{\textit{k}})$ function provided with the \textit{PY-BerryAHC} code further enhance its capability in exploring topological materials.

\bibliographystyle{apsrev4-2}
\bibliography{references} 

\end{document}